\documentclass [9pt, twocolumn, conference] {IEEEtran}
\usepackage{fancyhdr}

\renewcommand{\headrulewidth}{0pt}
\usepackage[perpage,symbol]{footmisc}
\usepackage {mathptm}
\usepackage[english]{babel}
\usepackage {graphicx}
\usepackage {subfig}
\usepackage {multirow}
\usepackage{epstopdf}
\usepackage{setspace}
\usepackage{booktabs,caption,fixltx2e}
\usepackage[flushleft]{threeparttable}
\usepackage{algorithm,algpseudocode}
\usepackage{amsmath}
\algrenewcomment[1]{\bgroup\hfill//~#1\egroup}
\usepackage{color}
\usepackage{setspace} 
\begin{document}

\thispagestyle{plain}
\pagestyle{plain}
\title{Hardware Trojan Detection through Information Flow \\ Security Verification} 

\author{\IEEEauthorblockN{Adib Nahiyan, Mehdi Sadi, Rahul Vittal, Gustavo Contreras, Domenic Forte and Mark Tehranipoor}
\IEEEauthorblockA{Department of Electrical and Computer Engineering, University of Florida, Gainesville, Florida}
{\normalsize Email: \{adib1991, mehdi.sadi,  rahulvittal, gustavokcm\}@ufl.edu}, \{dforte, tehranipoor\}@ece.ufl.edu}

\maketitle

\begin{abstract}

Semiconductor design houses are increasingly becoming dependent on third party vendors to procure intellectual property (IP) and meet time-to-market constraints. However, these third party IPs cannot be trusted as hardware Trojans can be maliciously inserted into them by untrusted vendors. While different approaches have been proposed to detect Trojans in third party IPs, their limitations have not been extensively studied. In this paper, we analyze the limitations of the state-of-the-art Trojan detection techniques and demonstrate with experimental results how to defeat these detection mechanisms. We then propose a Trojan detection framework based on information flow security (IFS) verification. Our framework detects violation of IFS policies caused by Trojans without the need of white-box knowledge of the IP. We experimentally validate the efficacy of our proposed technique by accurately identifying Trojans in the trust-hub benchmarks. We also demonstrate that our technique does not share the limitations of the previously proposed Trojan detection techniques.
 
\end{abstract}

%----------------------- This is special for ITC --------------------------
% YOur copyright string might be differfent

\thispagestyle{fancy}
\fancyhead{}
\renewcommand{\headrulewidth}{0pt}
\fancyhf{}
\fancyfoot[L]{\large Paper 6.1 \\ 978-1-5386-3413-4/17/\$31.00 \copyright2017 IEEE}
\fancyfoot[C]{\large INTERNATIONAL TEST CONFERENCE}
\fancyfoot[R]{\large \thepage}

%----------------------- This is special for ITC --------------------------

\section{Introduction}

Today's system-on-chips (SoCs) usually contain tens of IP cores (digital and analog) performing various functions. To lower research and development (R\&D) cost and speed up the development cycle, the SoC design houses typically purchase most of the IP cores from third-party (3P) vendors \cite{xiao_2016}. This raises a major concern toward the trustworthiness of 3PIPs because 3PIP vendors can insert malicious components (known as hardware Trojans) in their IPs \cite{adib_book_chapter}. This issue has gained significant attention as Trojans inserted by a 3PIP vendor can create backdoors in the design through which sensitive information can be leaked and other possible attacks (e.g., denial of service, reduction in reliability) can be performed \cite{Tehranipoor_10}. 

Detection of Trojans in 3PIPs is challenging as there is no golden version against which to compare a given IP core during verification. In theory, an effective way to detect a Trojan in an IP core is to activate the Trojan and observe its effects, but a Trojan's type, size, and location are unknown, and its activation condition is most likely a rare event. A Trojan can be, therefore, well hidden during the normal functional operation of the 3PIP supplied as register transfer level (RTL) code. A large industrial-scale IP core can include thousands of lines of code. Identifying the few lines of RTL code in an IP core that represents a Trojan is an extremely challenging task \cite{Tehranipoor_13}.

\begin{figure*}[t]
\centering
\includegraphics[width=.9\textwidth]{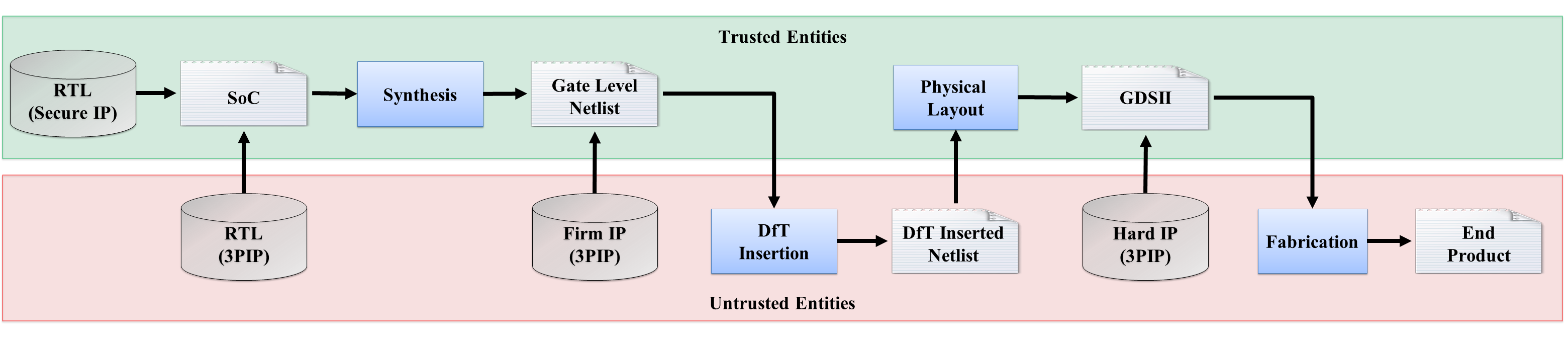}
\caption {System-on-chip (SoC) design flow}
\label{fig:SoC}
\end{figure*}

Different approaches have been proposed to validate that an IP does not perform any malicious function, i.e., an IP does not contain any Trojan. Existing Trojan detection techniques can be broadly classified into structural and functional analysis, logic testing, formal verification, information flow tracking and runtime validation. Structural analysis employs quantitative metrics to mark signals or gates with low activation probability as suspicious signals that may be a part of a Trojan \cite{Salmani2013} \cite{Salmani2013b}. Hicks et al. \cite{UCI} have proposed a technique named unused circuit identification (UCI) to find the lines of RTL code that have not been executed during simulation. These unused lines of codes can be considered to be part of a malicious circuit. However, these techniques do not guarantee Trojan detection and Sturton et al. \cite{defeat_UCI} have demonstrated that hardware Trojans can be designed to defeat UCI technique.

Functional analysis based approaches \cite{FANCI} \cite{VeriTrust} rely on functional simulation to find suspicious regions of an IP which have similar characteristics of a hardware Trojan. Waksman et al. \cite{FANCI} have proposed a technique named Functional Analysis for Nearly-unused Circuit Identification (FANCI) which flags nets having weak input-to-output dependency as suspicious. Zhang et al. \cite{VeriTrust} have proposed a technique called VeriTrust to identify nets that are not driven by functional inputs as  potential trigger inputs of a hardware Trojan. However, in \cite{DeTrust} authors have shown that a Trojan can be designed to bypass both of these detection techniques. Farahmandi et al. \cite{Farimah1} have proposed a Trojan detection technique based on symbolic algebra. However, this technique requires a golden reference of the IP which may not be available when it is procured from 3P vendors. 

Some recent works have utilized formal methods and information flow tracking (IFT) based techniques to detect Trojans. Rajendran et al. \cite{JV_1} have proposed a technique to formally verify malicious modification of critical data in 3PIP by hardware Trojans. Another similar approach has been proposed in \cite{JV_2} which formally verifies unauthorized information leakage in 3PIPs. In \cite{Jin1} \cite{Farimah2}, authors have proposed the concept of proof-carrying code (PCC) to formally validate the security related properties of an IP. Commercial tools, e.g., Jasper security path verification tool \cite{Jasper} can identify information leakage caused by design bugs and/or hardware Trojans. Wei et al. \cite{GLIFT} have used information flow tracking to detect Trojans in 3PIP cores. These state-of-the-art techniques have been shown to be effective in detecting various types of Trojans. However, all of these techniques share some inherent limitations which can be exploited to bypass their detection capabilities. 

In this paper, we present a comprehensive analysis of IFT/formal Trojan detection techniques. Also, we propose a novel framework to detect Trojans in 3PIPs addressing the shortcomings of previously proposed techniques. Our proposed framework is based on information flow security (IFS) verification which detects violation of IFS policies due to Trojans. Once we detect the presence of a Trojan, we extract its triggering condition. In this paper, we focus on utilizing our framework for Trojan detection. However, it can also be used to detect IFS violations unintentionally introduced by design mistakes or by CAD tools \cite{DSeRC} \cite{DSeRC_book}. We summarize our contribution as follows:

\begin{itemize}
\item We analyze the limitations of the recent Trojan detection techniques for 3PIP (\cite{JV_1} \cite{JV_2} \cite{Jasper} \cite{Jin1}) and demonstrate how these limitations can be exploited to bypass their detection capabilities.
\item We propose a IFS verification framework which detects violations of confidentiality and integrity policies caused by Trojan. The novelty of this approach is that it models an asset (e.g., a net carrying a secret key) as a fault and leverages the automatic test pattern generation (ATPG) algorithm to detect the propagation of the asset.
\item We propose a partial-scan ATPG technique to identify the observe/control points through/from which an asset can be leaked/influenced. Traditional full-scan ATPG and full-sequential ATPG cannot be used for this purpose. 
\item We propose a technique to differentiate between a valid asset propagation path and a Trojan payload path.
\item Our proposed technique can detect Trojan inserted by 3P vendors, e.g., design for test (DfT) vendors. This differentiates our approach from previously proposed techniques which cannot work with DfT inserted netlist. 
\item We experimentally validate the efficacy of our proposed technique using 18 benchmarks from the trust-hub \cite{Trusthub}.

\end{itemize}

The rest of the paper is organized as follows. In the next section, we give the necessary background on SoC design flow and discuss our threat model. In Section \ref{sec:Previous}, we analyze the limitations of the previously proposed Trojan detection technique and demonstrate with experimental results how to exploit them. Section \ref{sec:IFS} discusses our proposed framework which consists of confidentiality verification, integrity verification, and trigger condition extraction. We present our results in Section \ref{sec:results}. We discuss the limitations of our proposed approach in Section \ref{sec:discussion}. We conclude the paper with Section \ref{sec:concl}.

%----------------------- This is special for ITC --------------------------
% YOur copyright string might be differfent

\thispagestyle{fancy}
\fancyhead{}
\renewcommand{\headrulewidth}{0pt}
\fancyhf{}
\fancyfoot[L]{\large Paper 6.1}
\fancyfoot[C]{\large INTERNATIONAL TEST CONFERENCE}
\fancyfoot[R]{\large \thepage}

%----------------------- This is special for ITC --------------------------

\section{Preliminaries and Definitions}
\label{sec:Preliminaries}

\subsection{SoC Design Flow}
\label{sec:Soc}

Figure \ref{fig:SoC} shows a typical SoC design flow. Design specification by the SoC integrator is the first step. The SoC integrator then identifies a list of IPs necessary to implement the given specification. These IP cores are either developed in-house or purchased from 3PIP vendors. If purchased, these 3PIP cores can be procured in the following three forms~\cite{VLSI}, 

\begin{itemize}
	\item Soft IP cores are delivered as synthesizable register transfer level (RTL) written in hardware description language (HDL) such as Verilog or VHDL.
	
	\item Firm IP cores are delivered as gate level implementation of the IP, possibly using a generic library.

	\item Hard IP cores are delivered as GDSII representations of a fully placed and routed design.

\end{itemize}

After developing/procuring all the necessary soft IPs, the SoC design house integrates them to generate RTL specification of the whole SoC. The RTL design goes through extensive functional testing to verify the functional correctness of the SoC and also to find any design bugs. SoC integrator then synthesizes the RTL description into a gate-level netlist based on a target technology library. They may also integrate firm IP cores from a vendor into this netlist. The SoC integrator then integrates DfT structures to improve the design's testability. However, in many cases, the DfT insertion is outsourced to third party vendors who specialize in designing test and debug structures, e.g., built-in self-test (BIST), compression structures. In the next step, the gate-level netlist is translated into a physical layout design. It is also possible to import IP cores from vendors in GDSII layout file format and integrate them at this stage. After performing static timing analysis (STA) and power closure, SoC integrator generates the final layout in GDSII format and sends it out to a foundry for fabrication \cite{adib_book_chapter}.

Note that, we consider the entities inside the green box in Figure \ref{fig:SoC} as trusted while the entities inside the red box as rogue/untrusted.

\subsection{Threat Model}
\label{sec:threat}

In this section, we present how the potential adversaries can implant a hardware Trojan in a SoC design. We also briefly describe their objectives and their capabilities. Finally, we present the types of hardware Trojans covered by our threat model.

\subsubsection{Potential Adversaries}

We consider that all the 3PIPs, i.e., soft, firm and hard IPs as untrusted. However, in this work, we mainly focus on the soft and the firm IPs. The third party vendors involved in the SoC design flow (e.g., design for test (DfT) insertion) are considered untrusted as well. We consider the SoC integrator and the CAD tools as trusted entities. Although we do not trust the foundry, Trojan inserted by the foundry is out of the scope of this paper. Trojans inserted by foundry can only be detected by post-silicon verification techniques, whereas in this paper we focus on Trojan detection in pre-silicon design stage. 

We assume that the third party vendors will insert the Trojans while the SoC integrator will try to detect them. We consider the SoC integrator has black box knowledge of the IP purchased from the third party vendors. That is, the SoC integrator only has the knowledge of the high-level functionality of the IP. The 3PIP vendors have full control over their IP and can insert stealthy Trojans which would be extremely difficult, if not impossible, to detect using traditional test and verification techniques. The other important players in SoC design flow are the third party vendors, e.g., DfT vendors who have access to the whole SoC design and have the capability to incorporate stealthy malicious circuitry in the SoC. 

The objective of the adversaries is to implant Trojan in the SoC design through which sensitive information can be leaked and other possible attacks (e.g., denial of service, reduction in reliability) can be performed.

\subsubsection{Hardware Trojan Structure}
\label{sec:Trojan}

The basic structure of a hardware Trojan in a 3PIP can include two main parts, trigger, and payload. A Trojan trigger is an optional part that monitors various signals and/or a series of events in the SoC. Once the trigger detects an expected event or condition, the payload is activated to perform a malicious behavior. Typically, the trigger is expected to be activated under extremely rare conditions, so the payload remains inactive most of the time. When the payload is inactive, the SoC acts like a Trojan-free circuit, making it difficult to detect the Trojan \cite{adib_book_chapter}.

A Trojan can have a variety of possible payloads. In this paper, we focus on payloads which leak secret information (violation of confidentiality) and/or allows an adversary to gain unauthorized access to a privilege system (violation of integrity). Note that, most Trojan payloads will either violate the confidentiality and/or the integrity policies of the SoC. 

In this paper, we classify Trojan into two broad categories, 

\begin{itemize}
	\item \textbf{Trojan type I:} This type of Trojan delivers its payload through the valid asset propagation path. For example, RSA-T100 trust-hub Trojan \cite{Trusthub} leaks the private key through valid ciphertext output port. This type of Trojans generally creates a bypass path and allows an adversary to extract the asset. 
	
	\item \textbf{Trojan type II:} This type of Trojan delivers its payload through a malicious circuit which is not authorized to observe or control the asset. For example, AES-T100 trust-hub Trojan \cite{Trusthub} leaks the private key through a leakage circuit which is functionally isolated from the valid encryption logic.

\end{itemize}

The structure and the functionality of RSA-T100 and AES-T100 Trojans are discussed in details in Section \ref{sec:results}.

\begin{figure}[t]
\centering
\includegraphics[width=.35\textwidth]{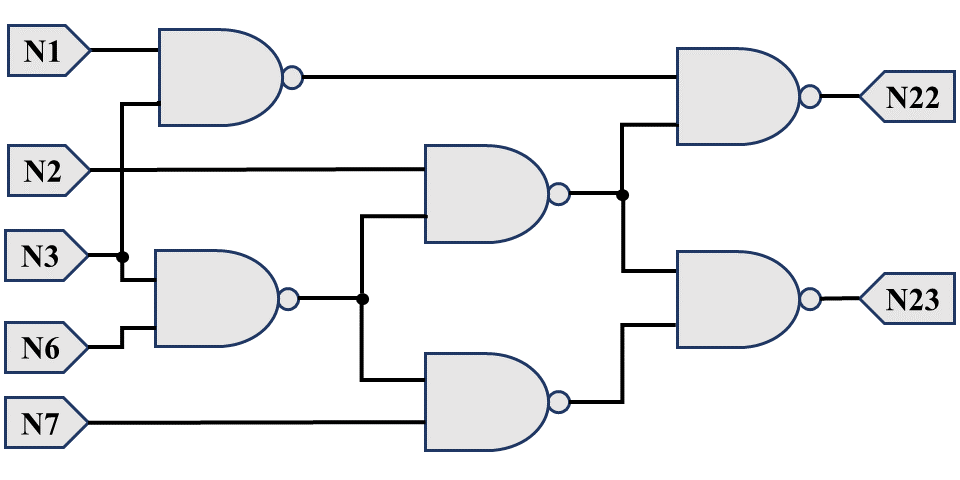}
\caption {C17 benchmark circuit}
\label{C17}
\end{figure}

\section{Previous Works and Their Limitations}
\label{sec:Previous}

Limitations of existing Trojan detection approaches, e.g., UCI \cite{UCI}, FANCI \cite{FANCI}, VeriTrust \cite{VeriTrust} have been discussed in details in \cite{defeat_UCI}, \cite{DeTrust}, \cite{JV_1}. In this paper, we analyze recent Trojan detection techniques which utilize formal methods and/or information flow tracking. 

\subsection{Formal Verification}
\label{sec:Formal}

Rajendran et al. \cite{JV_1} proposed a technique to formally verify malicious modification of critical data in 3PIP by hardware Trojans. Another similar approach has been proposed in \cite{JV_2} which formally verifies unauthorized information leakage in 3PIPs. Both techniques are based on bounded model checking (BMC) where the BMC checks for the property-``does critical information get corrupted?" and ``does the design leak any sensitive information?". The underlying concept is same for both techniques, and therefore, we will focus our analysis on \cite{JV_2}.

To check for information leakage caused by Trojan, authors in \cite{JV_2} used the following property, 
\vspace{-1ex}
\begin{equation}
P \models (s_0 == o) \vee (\neg s_0 == o),~~\forall~s_0 \in {0,1}
\label{eqn:formal1}
\end{equation}
where, $s_0$ is the individual bit of sensitive information (e.g., encryption key) and $\neg s_0$ is the inverted logical value of $s_0$. $o$ is any leakage point (e.g., output port). The authors claim that this property can detect if the sensitive information or a function of the sensitive information is leaked.

However, there exist many limitations of this technique, as discussed below:

\noindent
{\bf False Positive Results:}  The fundamental limitation of this technique is that it will produce a large number of false positive results. The reason is that the property only checks if the logical value of $s_0$ or its inverted value $\neg s_0$ is same as the logical value of $o$. However, the property does not guarantee if there exists an information flow from $s_0$ to $o$ causing the possibility of false positive results. We illustrate this limitation using the simple C17 benchmark circuit. 

Figure \ref{C17} shows the schematic of the C17 benchmark circuit. It is clear from Figure \ref{C17} input $N1$ is not in the fan-in cone of the output $N23$. In other words, there is no information flow from $N1$ to $N23$. Now we write the following assertion to prove the property of Equation \ref{eqn:formal1}.
\vspace{-1ex}
\begin{equation}
P \models assert~never~((N1 == N23) \vee (\neg N1 == N23))
\label{eqn:formal2}
\end{equation}

\begin{figure}[b]
\centering
\includegraphics[width=.35\textwidth]{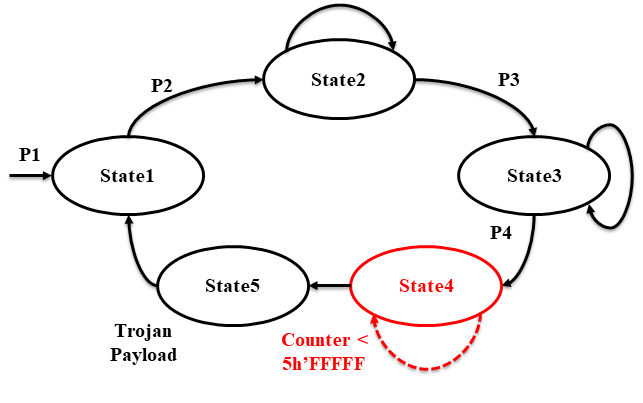}
\caption {Finite state machine of the modified triggering circuit of AES-T1100 Trojan}
\label{FSM}
\end{figure}

We use the inclusive formal verification (IFV) tool of Cadence \cite{cadence} to verify this property. The IFV tool returns that the assertion has failed which means that there exists information leakage from $N1$ to $N23$. This is a false positive result as there exists no information flow from $N1$ to $N23$. The reason for this false positive result is that the  IFV tool uses other input pins to make the logical value of $N23$ equal to $N1$. This result illustrates that the technique proposed in \cite{JV_1}, \cite{JV_2} would produce a large number of false positive results for any practical circuit.

\noindent
{\bf Limited Verification Capability:} Another fundamental limitation of any model checking technique is that a sequential circuit can be unrolled or verified only to a limited number of clock cycles, $T$. If the Trojan is designed in such a way so that its trigger gets activated after $T$, then the Trojan would evade detection. In \cite{JV_1}, \cite{JV_2} the authors acknowledged this limitation and proposed a trivial solution that the design needs to be reset once the number of clock cycles exceeds $T$. 

This proposed solution, however, does not provide adequate protection from Trojans. To demonstrate this, we take the AES-T1100 Trojan benchmark circuit \cite{Trusthub} from trust-hub. The triggering circuit of this Trojan can be illustrated as a finite state machine (FSM) consisting of four states (shown in black color in Figure \ref{FSM}). This Trojan is triggered when it observes four specific plaintexts ($P1$, $P2$, $P3$, $P4$) in the correct order. We use the property of Equation \ref{eqn:formal1} and use the IFV tool to detect the Trojan. The IFV tool easily detects the Trojan and also shows the triggering condition. Next, we modify the Trojan triggering circuit by incorporating one more state (State 4 in Figure \ref{FSM}) and a 20 bit counter. Once the Trojan observes the four specific triggering sequence, it starts the counter and waits for $2^{20}$ cycles before triggering its payload (State 5 in Figure \ref{FSM}). We now use the IFV tool again to detect the Trojan and the tool now fails to detect it. We also designed the Trojan counter and FSM trigger in such a way that these registers are not reset-able. Therefore, even when the design is reset, these registers store their current state and deliver its payload when the triggering condition is met. Our modification only introduces $0.4\%$ area overhead compared to the original AES-T1100 Trojan. Also, non-reset-able FFs are not uncommon, e.g., DFFX1 is a non-reset-able FF in the Synopsys standard cell library. Therefore, our proposed modification does not make the Trojan circuit distinguishable from the rest of the circuit. This experiment illustrates another crucial weakness of the formal methods in detecting Trojan. 

\begin{figure}[t]
\centering
\includegraphics[width=.45\textwidth]{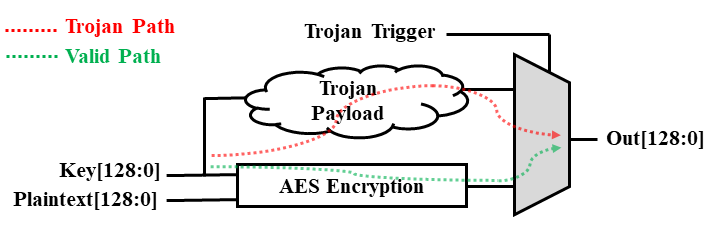}
\caption {AES encryption module with a hardware Trojan. The Trojan leaks the secret key through the ciphertext output when it is triggered.}
\label{GLIFT_Trojan}
\end{figure}

%----------------------- This is special for ITC --------------------------
% YOur copyright string might be differfent

\thispagestyle{fancy}
\fancyhead{}
\renewcommand{\headrulewidth}{0pt}
\fancyhf{}
\fancyfoot[L]{\large Paper 6.1}
\fancyfoot[C]{\large INTERNATIONAL TEST CONFERENCE}
\fancyfoot[R]{\large \thepage}

%----------------------- This is special for ITC --------------------------

\subsection{GLIFT-based Trojan Detection}
\label{sec:GLIFT}

GLIFT-based Trojan detection technique \cite{GLIFT} relies on gate-level information flow tracking (GLIFT) \cite{Hu_2014} to detect Trojan. To account for hardware specific information flow, GLIFT technique tracks information flow through Boolean gates. At gate level, all information flow appears at the most basic level of abstraction which allows detecting information flow that is inherently not visible at higher levels (software level). In GLIFT each data bit is associated with a taint bit and the propagation of data is monitored by tracking the tainted bits as they flow through Boolean gates. To track the propagation of taint bits, each standard cell gate is augmented with its corresponding tracking logic gate (referred to as GLIFT logic). However, GLIFT logic generation is a difficult problem due to its inherent complexity. Also GLIFT logic can produce false positive results as shown in \cite{Hu_2014}.

To detect Trojan using GLIFT, Wei et al. \cite{GLIFT} label the secret key as HIGH (tainted) and all the remaining inputs as LOW (untainted). Then they write formal properties to check if an unauthorized output port can have HIGH taint. In an encryption module any output port apart from the ciphertext outputs is considered as unauthorized output point as the secret key should not propagate to this point. The main advantage of GLIFT over formal verification based techniques, i.e., \cite{JV_1}, \cite{JV_2} is that formal methods only check functional properties, e.g., a signal $X$ can take a logic value of $Y$ \cite{GLIFT}. This does not reveal the actual information flow, raising the possibility of false positive results as shown in Section \ref{sec:Formal}. On the contrary, GLIFT can track the actual information flow. 

However, this technique also has some inherent limitations which can be exploited to bypass this technique. 

\noindent
{\bf Indistinguishable Valid and Malicious Paths:} GLIFT-based Trojan detection technique assumes that ``it is normal for the key to flow to the ciphertext output in a cryptographic function" \cite{GLIFT}. However, the proposed technique fails to take into account that an adversary can design a Trojan which leaks the key through ciphertext output. Figure \ref{GLIFT_Trojan} shows such a Trojan whose payload leak keys through a ciphertext output when a certain trigger condition is applied. The GLIFT-based Trojan detection technique would not be able to distinguish between the valid key propagation path and the key leakage path caused by the Trojan. This limitation raises the possibility of false negative results. 

\noindent
{\bf Taint Explosion:} Another major limitation of the GLIFT-based technique is that it cannot be applied to detect Trojans inserted by rogue DfT vendors. As discussed in Section \ref{sec:threat}, in many cases the DfT insertion is often outsourced to third party vendors who have access to the whole SoC design and have the capability to incorporate a Trojan. In a DfT inserted design all the FFs in a scan chain will be connected in series in scan mode. Now, if a secret information (taint value HIGH) is propagated to any of these FFs, then in scan mode all the FFs in the scan chain will be tainted as HIGH as well, causing a taint explosion. Therefore, GLIFT technique cannot be applied to detect Trojan in DfT inserted netlist. One may argue that it is possible to apply GLIFT in only functional mode by applying constraint in the scan enable signal. However, it has been demonstrated that it is possible to design Trojan that delivers its payload in scan mode, e.g., s35932-T100 Trojan in trust-hub \cite{Trusthub}. Such Trojans cannot be detected using GLIFT-based Trojan detection technique. 

Also, GLIFT technique relies on formal tools to detect Trojan and therefore, shares the same limitation of \textit{Limited Verification Capability} of formal methods.

\subsection{Jasper Security Path Verification}
\label{sec:Jasper}

With the growing importance of data security assurance, commercial tool vendors have started introducing security verification tools. JasperGold Security Path Verification (SPV) tool of Cadence \cite{Jasper} is a powerful formal verification tool that accepts a RTL code containing a specific secure area (memory or key location), and exhaustively proves that secure information: (i) cannot be read illegally, (ii) cannot be illegally overwritten

Jasper tool formally checks whether there exists any functional path from source signal $A$ to destination signal $B$. The tool first injects a unique tag called ``taint" on $A$ and utilizes a proprietary path sensitization technology to verify whether this unique tag can ever appear at $B$. Among many applications, Jasper tool can detect Trojans which cause information leakage. 

However, Jasper tool may not be effective in detecting certain types of Trojan. Jasper SPV tool checks whether there exists any functional path from asset to a destination signal. A SoC integrator having black-box knowledge of a 3PIP may not know which internal signal to check for that is associated with a Trojan. For example, AES-2000 or AES-T2100 \cite{Trusthub} Trojans direct the key to internal registers and rely on dynamic power consumption of these registers to leak the key. Such Trojans may not be detected using Jasper as a SoC integrator may not know which specific signals to look for to detect these Trojans.

We applied the Jasper security path verification on AES-T2000 and AES-T2100 Trojan benchmark circuits \cite{Trusthub} and analyzed for information leakage of a key bit to all output ports. However, we were unable to detect these Trojans as we do not know which specific internal signal to analyze for information leakage.

\begin{figure}[b]
\centering
\includegraphics[width=.45\textwidth]{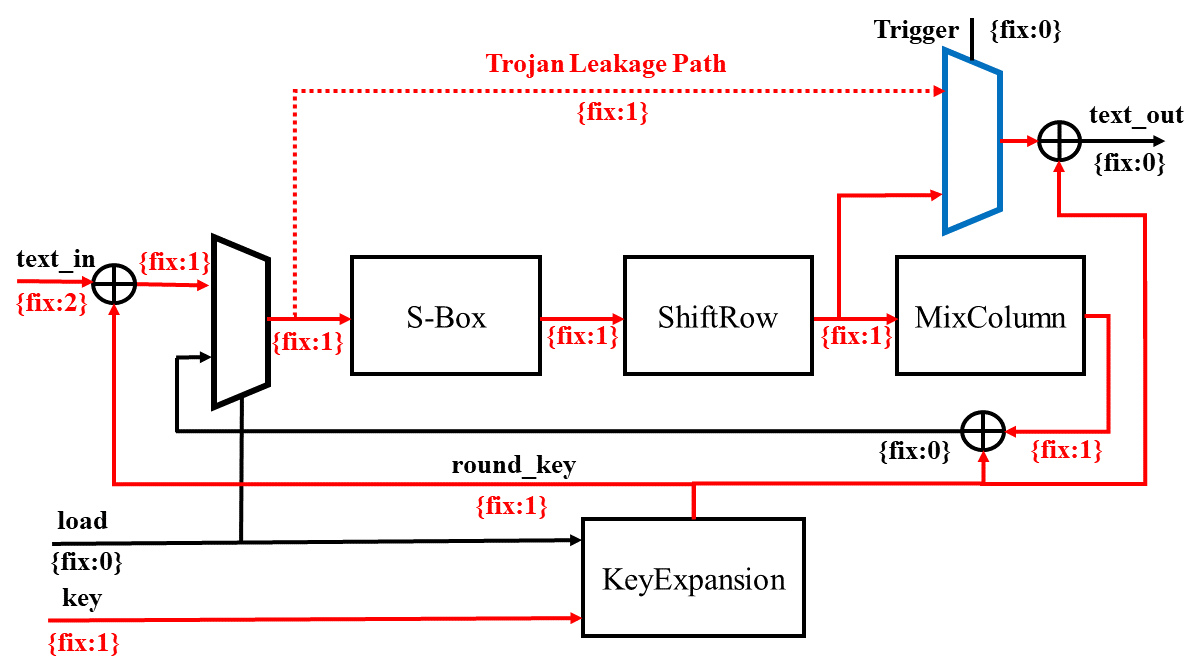}
\caption {Stable sensitivity status of AES encryption module. The red dotted line represents the Trojan leakage path and the blue colored mask is part of the Trojan payload.}
\label{PCC_AES}
\end{figure}

%----------------------- This is special for ITC --------------------------
% YOur copyright string might be differfent

\thispagestyle{fancy}
\fancyhead{}
\renewcommand{\headrulewidth}{0pt}
\fancyhf{}
\fancyfoot[L]{\large Paper 6.1}
\fancyfoot[C]{\large INTERNATIONAL TEST CONFERENCE}
\fancyfoot[R]{\large \thepage}

%----------------------- This is special for ITC --------------------------

\subsection{Trojan Detection by Signal Sensitivity Tracing}
\label{sec:PCC}

Jin et al. \cite{Jin1} have proposed the concept of proof-carrying code (PCC) to formally validate the security related properties of an IP and detect the possible Trojan presence. PCC methodology was proposed to perform data sensitivity tracing. The data sensitivity list contains the overall information of the distribution of the sensitive information across a design and this list can be evaluated to detect possible Trojan payload which leaks the sensitive information. The sensitivity of circuit signals which does not contain sensitive data are allocated with $0$ value while the signals containing sensitive data are allocated with positive integer values. A larger value indicates a higher level of the sensitivity of the signal and therefore, requires a higher level of protection. The signal transition value needs to be updated as the signal goes through different circuit operations. Their proposed technique uses a conservative model which allows only a few operations to downgrade the signal sensitivity list.

Figure \ref{PCC_AES} shows the stable signal sensitivity status of an AES encryption module. As shown in Figure \ref{PCC_AES}, the sensitivity value of ciphertext output is $\{fix:0\}$. The authors assume that a Trojan circuit which leaks sensitive information will cause sensitivity value of ciphertext output to be a positive integer instead of $\{fix:0\}$. 

\noindent
{\bf Bypassing Sensitivity Checking:} It is possible to design a Trojan which leaks the sensitive information without making the sensitivity value of ciphertext output a positive integer. An attacker can introduce a Trojan leakage path which bypasses the S-Box, ShiftRow and MixColumn operation, and leaks the intermediate state to the output (shown with a red dotted line in Figure \ref{PCC_AES}). Note that, this Trojan path does not affect the sensitivity value of ciphertext output. However, this Trojan leaks the intermediate state ($Plaintext \oplus Key$) when it is triggered allowing the attacker to recover the key from the known plaintext. This Trojan bypasses the proposed detection technique as it does not affect the sensitivity value of ciphertext output. We experimentally validate this limitation using the AES-T100 Trojan benchmark \cite{Trusthub} in Section \ref{Comparison}.

\begin{figure}[t]
\centering
\includegraphics[width=.5\textwidth]{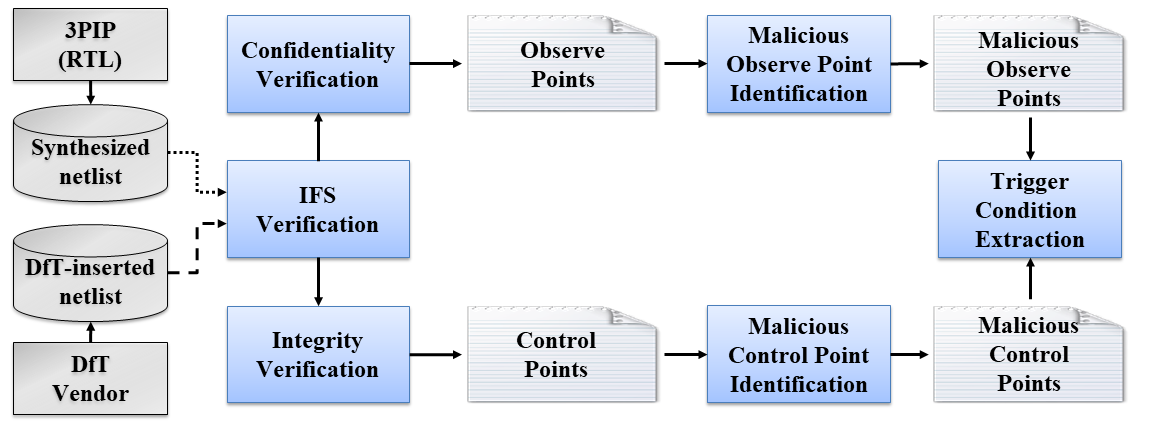}
\caption {Overview of our proposed IFS verification framework for Trojan detection.}
\label{IFS}
\end{figure}

%----------------------- This is special for ITC --------------------------
% YOur copyright string might be differfent

\thispagestyle{fancy}
\fancyhead{}
\renewcommand{\headrulewidth}{0pt}
\fancyhf{}
\fancyfoot[L]{\large Paper 6.1}
\fancyfoot[C]{\large INTERNATIONAL TEST CONFERENCE}
\fancyfoot[R]{\large \thepage}

%----------------------- This is special for ITC --------------------------

\section{Trojan Detection through IFS Verification}
\label{sec:IFS}

In Section \ref{sec:Previous}, we have shown that recent Trojan detection techniques suffer from large false positive results and limited verification capability. Also, these techniques cannot distinguish between a valid and a Trojan propagation path and is also incapable of detecting Trojans in DfT inserted netlist. We propose a novel framework to detect Trojan in 3PIPs which addresses these limitations.

Our proposed Trojan detection framework is based on information flow security (IFS) verification which detects violation of IFS policies due to Trojans. This framework can be applied to synthesized gate-level netlist of a design. To detect Trojan using IFS framework, a SoC integrator will first synthesize the 3PIP to get the gate-level netlist and then apply the framework. Another advantage of IFS verification technique is that it can also detect Trojan in a DfT inserted netlist unlike previously proposed techniques (see Section \ref{sec:GLIFT}). Therefore, the proposed technique can be applied to detect Trojan inserted by DfT vendors. Our Trojan detection framework is based on the observation that a Trojan, however, small, will alter the normal information flow of a design and thereby violate information flow security (IFS) policies. The challenge here is to detect IFS violation caused by Trojan without using any golden reference model. 

Our IFS verification framework is based on a novel concept of modeling an asset (e.g., a net carrying a secret key) as a stuck-at-0 and stuck-at-1 fault and leveraging the automatic test pattern generation (ATPG) algorithm to detect that faults. A successful detection of faults means that the logical value of the asset carrying net can be observed through the observe points or logical value of the asset can be controlled by the control points. In other words, there exists information flow from asset to observe points or from control points to asset. Here, the observe points refer to any primary or pseudo-primary (scan FFs) outputs that can be used to observe internal signals and the control points refer to the primary or pseudo-primary (scan FFs) inputs that can be used to control internal circuit signals.

For our IFS verification framework, we need to identify all observe points through which an asset can be observed and identify all control points from which an asset can be controlled. There are certain challenges of using conventional full-scan and full-sequential ATPG analysis for IFS verification framework. In full-scan ATPG, we can detect the asset propagation only to the first level FFs. Asset propagation to the subsequent level of FFs cannot be performed using full-scan ATPG \cite{gus}. This presents a major limitation because the observe points associated with a Trojan which is not located in the first level FFs, will not be detected by full-scan ATPG. In full-sequential ATPG method, a sequence of functional input stimulus is generated to activate and propagate a fault to an observe point. However, even for a simple stuck-at fault, the full-sequential ATPG needs to search through the space of all possible test vector sequences. Due to the high complexity of the full-sequential ATPG, it remains a challenging task to detect fault propagation in sequential circuits that do not incorporate any DfT scheme \cite{seq_ATPG}.

\begin{figure}[b]
\centering
\includegraphics[width=.45\textwidth]{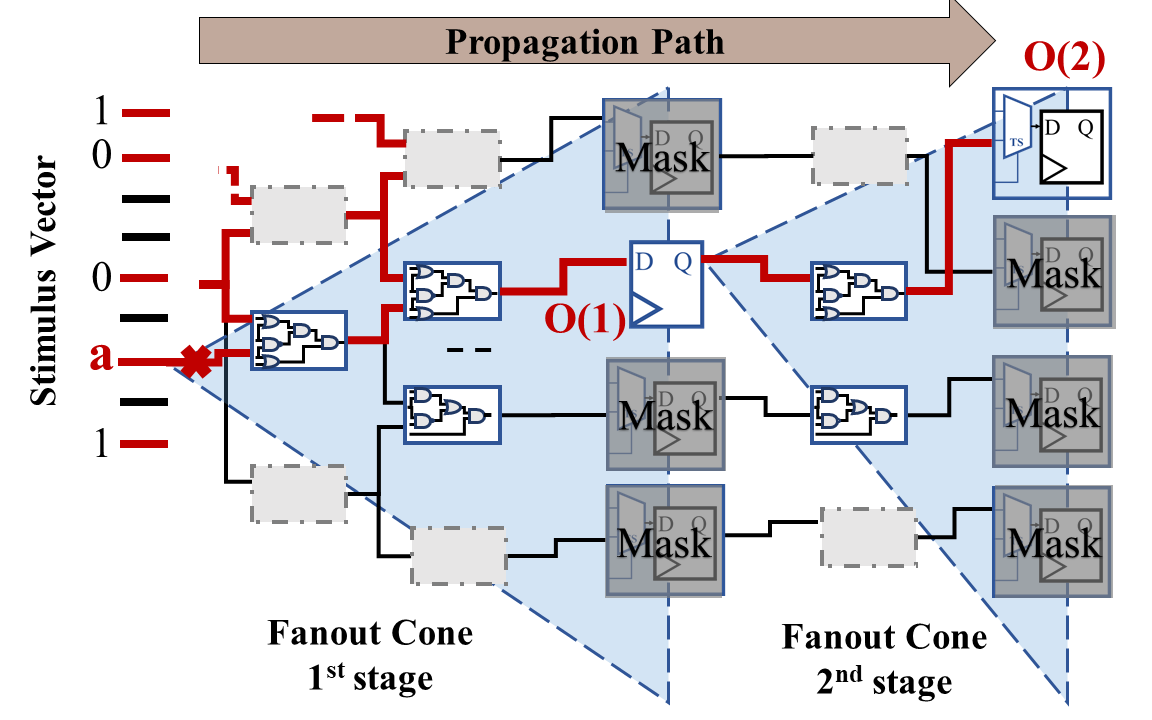}
\caption {Confidentiality verification algorithm utilizes partial-scan ATPG to identify the observe points through which an asset can be observed. The algorithm utilizes \textit{add\_scan\_ability}, masking and successive fan-out analysis. Here, $O(1)$ and $O(2)$ represent the first and second stage observe points, respectively.}
\label{confidentiality}
\end{figure}

We overcame the limitations of full-scan and full-sequential ATPG with our proposed \textit{partial-scan ATPG} technique which is capable of identifying the observe/control points of an asset. In a partial-scan design, the scan chains contain some, but not all, of the sequential cells in the design. Traditionally, partial-scan is used to minimize area overhead (caused by DfT structure) while attaining targeted test coverage. However, in our IFS framework, we have used the partial-scan ATPG to identify the observe points through which an asset can be observed and identify the control points from which an asset can be controlled.  Note that, the partial-scan ATPG technique is only used for our IFS verification purpose. Once the verification is performed, the design can be transformed into a full-scan design to improve test coverage. 

Figure \ref{IFS} shows the overview of the IFS verification framework for Trojan detection. This technique can detect Trojan inserted by 3PIP and DfT vendors. The IFS verification can be further classified to \textit{Confidentiality Verification} and \textit{Integrity Verification}. In \textit{Confidentiality Verification}, we analyze all the observe points through which an asset can be observed and identify if the asset can be observed through any unauthorized/malicious observe points. Similarly, in  \textit{Integrity Verification}, we identify if an asset can be controlled by any unauthorized/malicious control points. We distinguish between authorized and unauthorized observe/control points using our proposed \textit{Malicious Observe/Control Point Identification} technique. This technique allows us to detect Trojans without the need of white box knowledge of the IP and golden reference model. Once we have identified any IFS verification failure, we apply the \textit{Trigger Condition Extraction} technique to extract the triggering condition of the Trojan.

It is important to note that IFS framework utilizes existing test tools such as Tetramax from Synopsis, Fastscan from Mentor Graphics, or Encounter Test from Cadence which is commercially available and widely used by industry and academia. In this paper, we have used Tetramax \cite{synopsys_tetra} along with our developed tcl code to implement the IFS verification framework. 

In the subsequent subsections, we describe the various steps of our proposed IFS verification framework in details. 

\begin{algorithm}[t]
\caption{Confidentiality Verification}
\label{alg:confidentiality}
\begin{algorithmic}[1]
\footnotesize
	\Procedure{Confidentiality Verification}{}\\
	\textbf{Input:} Asset, gate-level netlist, technology library\\
    \textbf{Output:} Observe points, asset propagation path, stimulus vector
	\State $FF\_level \gets 1$ 
    \State $add\_scan\_ability(all\_register)$
		
	\ForAll{$a \in Asset$} 
	
		\State $FanoutFinal \gets fanout(a, endpoints\_only)$
		\State $mask\_register(FanoutFinal)$
		\While{$1$} 
		
		  \ForAll{$FO \in FanoutFinal$} 
		  
			\State $unmask\_register(FO)$
			\State $add\_Faults(a, ``stuck-at")$ 
			\State $run\_sequential\_ATPG$
			\State $analyze\_Faults(a, ``stuck-at\ 0")$ 
			\State $analyze\_Faults(a, ``stuck-at\ 1")$ 
			 
			 \If {$detected~faults~>~1$}
				\State $append~ObsPoints \gets FO$
				\State $Report:~propagation path(a, FO), control sequence$
				\State $append~FanoutTemp \gets fanout(FO, endpoints\_only)$
				\State $remove\_scan\_ability(FO)$
				\State $mask\_register(FO)$
			\EndIf
			
		  \EndFor
		  
		  \If {$FanoutTemp\ is\ not\ empty$} 
					\State $FanoutFinal\gets FanoutTemp$
					\State $FF\_level ++$
				\Else
					\State $break$
				\EndIf
			
		\EndWhile 
	\EndFor
	\EndProcedure
\end{algorithmic}
\end{algorithm}

%----------------------- This is special for ITC --------------------------
% YOur copyright string might be differfent

\thispagestyle{fancy}
\fancyhead{}
\renewcommand{\headrulewidth}{0pt}
\fancyhf{}
\fancyfoot[L]{\large Paper 6.1}
\fancyfoot[C]{\large INTERNATIONAL TEST CONFERENCE}
\fancyfoot[R]{\large \thepage}

%----------------------- This is special for ITC --------------------------

\subsection{Confidentiality Verification}
\label{sec:Confidentiality}

Confidentiality policy ensures that the information from a classified system never leaks to an unclassified one. For example, a secret key for data encryption should never flow to an 
unclassified domain. A violation of the confidentiality policy indicates that an asset can be leaked through an observe point which is accessible to an attacker. In \textit{Confidentiality Verification}, we first identify all the observe points through which an asset can be observed and analyze if the asset can be observed through any unauthorized observe points.

Our proposed confidentiality verification technique is presented in Algorithm \ref{alg:confidentiality} and shown in Figure \ref{confidentiality}. The algorithm, first takes an asset (name of the port or net where the asset is located), the gate-level netlist of the design and the technology library (required for ATPG analysis) as inputs (Line 2). Then, the algorithm adds scan capability to all the FFs in the design to make them controllable and observable (see line 5). Here, we use the ``What If" analysis capabilities provided by the Tetramax software which allows one to add and/or remove FFs from scan chain. This feature allows us to perform partial-scan analysis dynamically without requiring to re-synthesize the netlist. Now, for each asset $a \in Asset$, the algorithm finds the observe points (primary output or scan FFs) in the fan-out cone of $a$ (Line 7). The algorithm then adds capture masks in these FFs (Line 8) so asset propagation to any observe point can be individually tracked. Next, for each observe points $FO$, the algorithm removes the mask from (Line 11) so that propagation of $a$ to $FO$ can be detected. Line 12 of Algorithm \ref{alg:confidentiality} adds asset $a$ as the only stuck-at fault in the design. Lines 13-15 use ATPG algorithm in the sequential mode to find paths to propagate $a=0$ and $a=1$ to observe point $FO$. If both, $a=0$ and $a=1$ is detected from $FO$ (Line 16), then there exists an information flow from $a$ to $FO$ and the algorithm marks $FO$ as an observe point and reports the propagation path from $a$ to $FO$ as well as the control sequence or stimulus required for the asset propagation (Line 17-18). Next, the algorithm finds the next level observe points by analyzing the fan-out cone of $FO$ (Line 19). Also, the scan ability is removed from $FO$ and thereby allowing the asset to propagate to next level observe points through $FO$ using sequential ATPG (Line 20). This process continues until a level of observe point is reached where all the observe points are primary outputs or the sequential ATPG algorithm cannot generate patterns to propagate the fault to observe point (Line 24-28). The output of the algorithm is the list of observe points, and the propagation path along with the stimulus vector for asset propagation for each observe points. 

\begin{figure}[b]
\centering
\includegraphics[width=.45\textwidth]{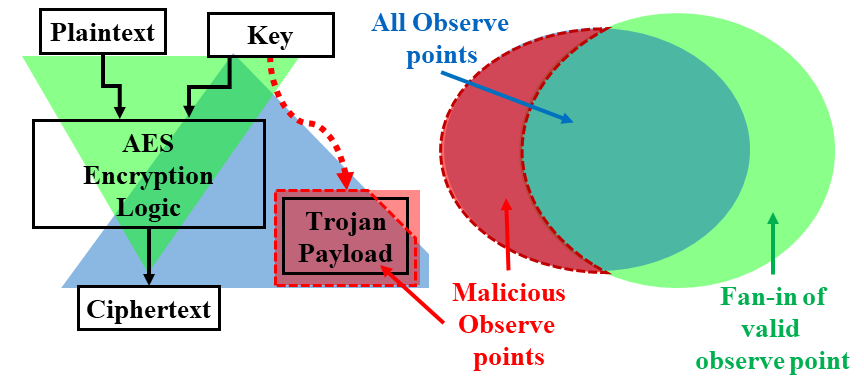}
\caption {Overview of our proposed intersect analysis. The blue region represents the all observe point where an Asset can propagate to and the green region represents all the fan-in elements of valid observe points (Ciphertext output port). The red dotted region represents the malicious observe points (Type II Trojan).}
\label{fig:intersect}
\end{figure}

\noindent
{\bf Malicious Observe Point Identification:} Algorithm \ref{alg:confidentiality} identifies the observe points through which an asset can be observed. However, in order to detect the presence of a Trojan, we need to distinguish between valid/authorized and malicious observe points. 

In Section \ref{sec:Trojan} we have broadly classified Trojans into two types. The \textit{Trojan Type I} leaks the asset through valid observe points by creating bypass paths for asset leakage. This type of Trojan can be detected by analyzing the asset propagation path reported by Algorithm \ref{alg:confidentiality}. The propagation path report contains propagation depth which represents the total number of gates an asset propagates through before reaching the observe points. The propagation depth for the Trojan bypass path will be much less than the propagation depth of valid propagation path. Here, one can argue that an attacker can add redundant logic to make the propagation depth of leakage path equal to the valid path. However, our technique is applied to the synthesized netlist and the synthesis process will automatically remove any redundant logic. Also, by analyzing the propagation path, we can retrieve which type of operations (e.g., and, or, xor) an asset goes through before propagating to an observe point. A SoC designer having the high-level knowledge of the 3PIP can also use this information to identify \textit{Type I} Trojans.

Now, \textit{Trojan Type II} leaks the asset through a leakage circuit which is functionally separated from the logic which is authorized to handle the asset. We distinguish the observe points associated with the leakage circuit by our proposed \textit{Intersect Analysis}. In this analysis, we record all the fan-in elements of the valid observe points. For example, in an encryption module ciphertext output ports are the valid observe points. If we find any FF which can observe the asset but is not located in the fan-in cone of the ciphertext output ports then we can conclude that the key can be leaked to an observe point which is not part of the encryption logic. In other words, there exists a violation of confidentiality policy in the given design. Figure \ref{fig:intersect} shows the \textit{Intersect Analysis}.

\begin{algorithm}[t]
\caption{Algorithm for Integrity Verification}
\label{alg:integrity}
\begin{algorithmic}[1]
\footnotesize
	\Procedure{Integrity Verification}{}\\
	\textbf{Input:} Asset, gate-level netlist, technology library\\
    \textbf{Output:} Control points, asset activation path, stimulus vector
	\State $FF\_level \gets 1$ 
    \State $add\_scan\_ability(all\_register)$
		
	\ForAll{$a \in Asset$} 
	
		\State $FaninFinal \gets fanin(a, startpoints\_only)$
		
		\While{$1$} 
		
			\State $add\_Faults(a, ``stuck-at")$ 
			\State $run\_sequential\_ATPG$
			\State $analyze\_Faults(a, ``stuck-at\ 0")$ 
			\State $analyze\_Faults(a, ``stuck-at\ 1")$ 
			 
			 \If {$detected~faults~>~1$}
				  \ForAll{$FI \in FaninFinal$} 
				     \State $append~ControlPoints \gets FI$
				     \State $Report:~ControlPath(a), control~sequence$
				     \State $append~FaninTemp \gets fanin(FI, startpoints\_only)$
				     \State $remove\_scan\_ability(FI)$
				  \EndFor
			\EndIf
					  
		  \If {$FaninTemp\ is\ not\ empty$} 
					\State $FaninFinal\gets FaninTemp$
					\State $FF\_level ++$
				\Else
					\State $break$
				\EndIf
			
		\EndWhile 
	\EndFor
	\EndProcedure
\end{algorithmic}
\end{algorithm}

%----------------------- This is special for ITC --------------------------
% YOur copyright string might be differfent

\thispagestyle{fancy}
\fancyhead{}
\renewcommand{\headrulewidth}{0pt}
\fancyhf{}
\fancyfoot[L]{\large Paper 6.1}
\fancyfoot[C]{\large INTERNATIONAL TEST CONFERENCE}
\fancyfoot[R]{\large \thepage}

%----------------------- This is special for ITC --------------------------

\subsection{Integrity Verification}
\label{sec:Integrity}

Integrity policy ensures that an untrusted system should never influence a trusted one. For example, an untrusted control point (e.g., a register), should never be able to influence a control pin of a trusted system. A violation of the integrity policy indicates that an asset can be influenced by control points which are accessible to an attacker. In \textit{Integrity Verification}, we first identify the control points through which an asset can be controlled and analyze if the asset can be influenced by any unauthorized control points.

\begin{figure}[b]
\centering
\includegraphics[width=.4\textwidth]{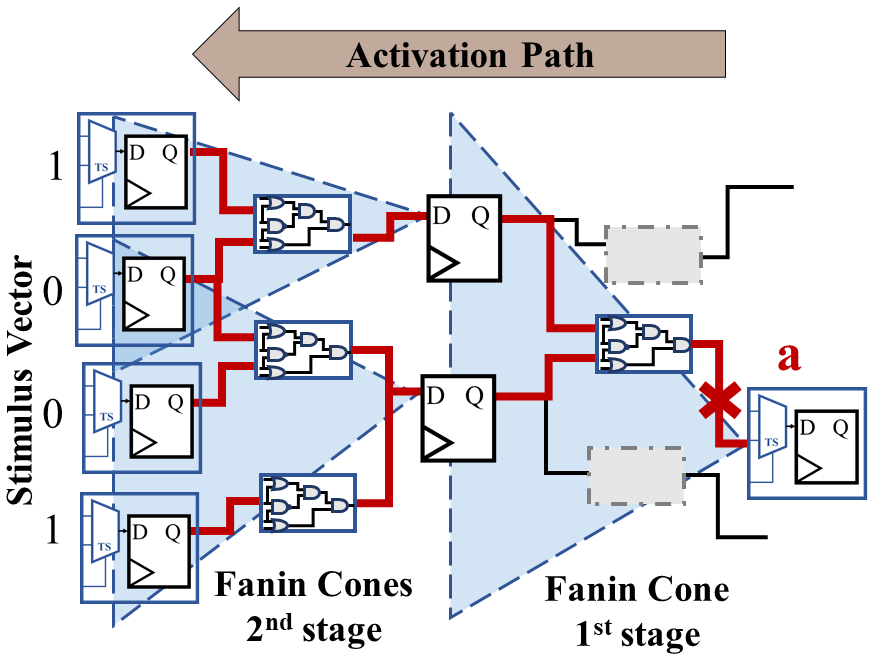}
\caption {Integrity verification algorithm utilizes partial-scan ATPG to identify all control points from which an asset can be influenced. The algorithm utilizes \textit{add\_scan\_ability}, masking and successive fan-in analysis.}
\label{integrity}
\end{figure}

Our proposed integrity verification technique is presented in Algorithm \ref{alg:integrity} and shown in Figure \ref{integrity}. The algorithm, first takes an asset (name of the register where the asset is located), the gate-level netlist of the design and the technology library (required for ATPG analysis) as inputs (Line 2). Then, the algorithm adds scan capability to all the FFs in the design to make them controllable and observable (see line 5). Now, for each asset $a \in Asset$, the algorithm finds the control points ($FaninFinal$) in the fan-in cone of $a$ (Line 7). Line 9 of Algorithm \ref{alg:integrity} adds asset $a$ as the only stuck-at fault in the design. Lines 9-12 use ATPG algorithms in the sequential mode to activate the fault. A successful detection of the faults indicates that there exists information flow $FaninFinal$ to $a$ (Line 13). Now for each $FI \in FaninFinal$, Algorithm \ref{alg:integrity} marks the $FI$ list as a control point and reports the activation path from $FI$ to $a$ as well as the control sequence or stimulus required to activate the faults (Line 15-16). Next, the algorithm finds the previous level control points by analyzing the fan-in cone of $FI$ (Line 17). Also, the scan ability is removed from $FI$ and thereby allowing the asset to be controlled by previous level control points (Line 18). This process continues until a level is reached where all the control points are primary inputs or the sequential ATPG algorithm cannot generate patterns to activate the faults from the control points (Line 21-25). The output of the algorithm is the list of control points and the activation path along with the stimulus vector for asset activation.

\noindent
{\bf Malicious Control Point Identification:} Algorithm \ref{alg:integrity} identifies all the control points which can influence an asset. We use similar techniques discussed in Section \ref{sec:Confidentiality} to distinguish between valid/ authorized and malicious control points. 

We detect the \textit{Trojan Type I} by analyzing the depth of asset activation path. The depth for the Trojan path will be much less than the depth of valid asset activation path. We detect the \textit{Trojan Type II} using our proposed \textit{Intersect Analysis}. Here, we record all the fan-out elements of the valid control points. For example, if we consider the program counter (PC) of a microprocessor as an asset then the valid control point for PC will be the pipeline register in the instruction decode stage. If we find any FF which can control the asset but is not located in the fan-out of PC then we can conclude that there exists a violation of integrity policy.

\begin{table*} [t]
 \centering
\caption{Results for Confidentiality Verification.}
\vspace{-1ex}
\begin{center}
\scriptsize
\centering
 %   \scriptsize\addtolength{\tabcolsep}{-4pt}
\begin{tabular}{|c|c|c|c|c|c|c|}
\hline
%\multirow{2}{*}{}  Benchmarks &  Trojan  & Trojan & Trojan &  \# of Observe  &  \# of Malicious  &  time  \\
% &  payload  & trigger & type   &  points  &  points  &  (s)  \\
%\hline
\textbf{Benchmarks} &  \textbf{Trojan payload} & \textbf{Trojan trigger} & \textbf{Trojan type} &  \textbf{\# of Observe points} &  \textbf{\# of Malicious points} &  \textbf{time(s)}  \\
\hline
AES-T100 &  Leaks the key through covert CDMA channel  & Always on & II &  42  &  16  &  251.5  \\
\hline
AES-T200 &  Leaks the key through covert CDMA channel  & Always on & II &  42  &  16  &  273.8  \\
\hline
AES-T700 &  Leaks the key through covert CDMA channel  & Specific plaintext & II &  42  &  16  &  277.1  \\
\hline
AES-T900 &  Leaks the key through covert CDMA channel  & Counter & II &  42  &  16  &  293.7  \\
\hline
AES-T1100 &  Leaks the key through covert CDMA channel  & Plaintext sequence (FSM) & II &  42  &  16  &  362.9  \\
\hline
AES-T2000 &  Leaks the key through PSC of shift register  & Specific plaintext & II &  35  &  1  &  240.5  \\
\hline
AES-T2100 &  Leaks the key through PSC of shift register  & Plaintext sequence (FSM) & II &  35  &  1  &  350.5  \\
\hline
RSA-T100 &  Leaks the key through ciphertext output  & Specific plaintext & I &  37  &  2  &  19.7 \\
\hline
RSA-T300 &  Leaks the key through ciphertext output  & Counter & I &  37  &  2  &  20.4 \\
\hline
DES-T100 &  Leaks the key through covert CDMA channel  & Always on & II &  28  &  16  &  79.7 \\
\hline
PRESENT-T100 &  Leaks the key through covert CDMA channel  & Always on & II &  19  &  16  &  1.53 \\
\hline
AES-T1100 (M) &  Leaks the key through covert CDMA channel  & Plaintext sequence (FSM) & II &  42  &  16  &  364.5  \\
\hline
\end{tabular}
\end{center}
\label{tab:confidentiality}
\end{table*}

%----------------------- This is special for ITC --------------------------
% YOur copyright string might be differfent

\thispagestyle{fancy}
\fancyhead{}
\renewcommand{\headrulewidth}{0pt}
\fancyhf{}
\fancyfoot[L]{\large Paper 6.1}
\fancyfoot[C]{\large INTERNATIONAL TEST CONFERENCE}
\fancyfoot[R]{\large \thepage}

%----------------------- This is special for ITC --------------------------

\subsection{Trigger Condition Extraction}
\label{sec:Trigger}

Our proposed \textit{Confidentiality} and \textit{Integrity Verification} coupled with \textit{Malicious Observe/Control Point Identification} identify IFS violation introduced by a Trojan. Now, we need to find the input sequence which triggers the Trojan. 

\textit{Confidentiality Verification} reports the stimulus vector which causes an asset to propagate to a malicious observe point whereas \textit{Integrity Verification} reports the stimulus vector which creates an activation path from a malicious control point to the asset. For some Trojans, the triggering sequence can be directly extracted from the stimulus vector. For example, in RSA-T100 \cite{Trusthub} monitors a particular plaintext to trigger its Trojan and AES-T900 \cite{Trusthub} triggers after a certain clock cycle. These triggering conditions are relatively simple and can be directly extracted from stimulus vector. However, there are Trojans e.g., AES-T1100 \cite{Trusthub} whose trigger circuit is composed of a finite state machine (FSM) and triggers when a sequence of plaintext is observed. The stimulus vector generated by IFS framework will report the register values ($Trig\_Conditon$) of Trojan trigger circuit which violates IFS policies. However, it will not report the functional input sequences which will cause the $Trig\_Conditon$. 

To extract the triggering condition for Trojans like AES-T1100, first, we need to determine if the registers associated with $Trig\_Conditon$ is part of an FSM. If the output of a register feeds back to its input through a series of combinational circuits, then it is a potential state register \cite{Shi}. The reason is that the next state of any FSM depends on the present state and therefore, any state register will have a feedback loop. Once we have identified the state registers, we use the FSM extraction technique proposed in \cite{nahiyan2016avfsm} to retrieve the functionality of the FSM. This technique determines the present states and input conditions which cause a transition to a particular state and repeat this process to extract the state transition graph (STG) of the overall FSM. For our case, we apply this technique to extract the present states and input conditions which cause a transition to a state represented by $Trig\_Conditon$. Then we repeat the process to generate the STG of the overall FSM. From the STG, we can extract the sequence of input patterns which triggers the Trojan.

One major advantage of this technique is that it reverse engineers the STG of the FSM and does not rely on functional simulation or unrolling of a sequential circuit. In Section \ref{sec:Formal} we have shown that formal method is only effective to a limited number of clock cycles and an attacker can easily exceed this limit by introducing additional states and counter. However, the FSM extraction technique is not limited by the number of clock cycles and is capable of detecting these types of Trojan, unlike the formal methods.  

\section{Results}
\label{sec:results}

\subsection{Experimental Setup}

We applied our IFS verification framework to detect Trojans in trust-hub \cite{Trusthub} benchmark circuits. We used the Trojan benchmarks which cause IFS violation. As mentioned in Section \ref{sec:IFS}, our proposed technique works with synthesized gate-level netlist. However, most trust-hub \cite{Trusthub} benchmark circuits are in RTL level. Therefore, we first synthesized the RTL benchmarks into gate-level netlist and then, apply our technique. All benchmarks were synthesized using Synopsys Design Compiler \cite{synopsys_dc} with Synopsys standard cell library. Note that, our technique works at gate level netlist, and therefore, is independent of the technology node. 

We used Tetramax \cite{synopsys_tetra} (Synopsys) along with our custom tcl scripts to perform the \textit{Confidentiality} and \textit{Integrity} verification. We implemented the state machine extraction technique \cite{nahiyan2016avfsm} using Matlab (for text parsing and analysis) and Tetramax (for pattern generation).

\begin{table*} [t]
 \centering
\caption{Results for Integrity Verification.}
\vspace{-1ex}
\begin{center}
\scriptsize
\centering
 %   \scriptsize\addtolength{\tabcolsep}{-4pt}
\begin{tabular}{|c|c|c|c|c|c|c|}
\hline
%\multirow{2}{*}{}  Benchmarks &  Trojan  & Trojan & Trojan &  \# of Observe  &  \# of Malicious  &  time  \\
% &  payload  & trigger & type   &  points  &  points  &  (s)  \\
%\hline
\textbf{Benchmarks} &  \textbf{Trojan payload} & \textbf{Trojan trigger} & \textbf{Trojan type} &  \textbf{\# of Control points} &  \textbf{\# of Malicious points} &  \textbf{time(s)}  \\
\hline
PIC-T100 &  Manipulates program execution flow & Counter & II &  17  &  13  &  0.358  \\
\hline
PIC-T200 &  Manipulates instruction register  & Counter & II &  41  &  14  &  100.5  \\
\hline
b19-T500 &  Manipulates instruction register  & FSM & II &  193  &  2  &  211.6  \\
\hline
RS232-T500 &  Manipulates a control signal  & Counter & II &  13  &  2  &  0.381  \\
\hline
s35932-T100 &  Manipulates scan mode  & Counter & II &  23  &  23  &  1.905  \\
\hline
RSA-T400 &  Replaces the key to leak plaintext  & Counter & II &  34  &  33  &  20.2  \\
\hline
\end{tabular}
\end{center}
\label{tab:integrity}
\end{table*}

%----------------------- This is special for ITC --------------------------
% YOur copyright string might be differfent

\thispagestyle{fancy}
\fancyhead{}
\renewcommand{\headrulewidth}{0pt}
\fancyhf{}
\fancyfoot[L]{\large Paper 6.1}
\fancyfoot[C]{\large INTERNATIONAL TEST CONFERENCE}
\fancyfoot[R]{\large \thepage}

%----------------------- This is special for ITC --------------------------

\subsection{Confidentiality Verification}

We applied our \textit{Confidentiality Verification} algorithm to detect twelve different Trojan benchmark circuits which violate the confidentiality policy. Table \ref{tab:confidentiality} summarizes our results. Column 1 represents which Trojan benchmark was used for our verification while column 2 and 3 represent the Trojan payload and Trojan trigger for each benchmark. Column 4 shows the Trojan types which was discussed in Section \ref{sec:Trojan}. In the Trojan benchmark circuits, all key bits are considered as assets. The analysis was performed for a single bit of the asset key for each implementation such as, for AES-T100, key[0] was used. Note that, we also experimented with other asset bits (e.g., key[1], key[2], etc.) and they produced identical results for the key bits which are leaked.

We first apply the \textit{Confidentiality Verification} to identify the observe points that can observe the asset. Column 5 shows the number of observe points for each asset bit. We then apply our \textit{Malicious Observe Point Identification} to identify the observe points associated with Trojans. Column 6 shows the malicious observe points identified by the proposed approach. Column 7 shows the required time in second for each analysis.

For all the twelve Trojan benchmark circuits, our algorithm was able to detect the confidentiality violation caused by the Trojan circuit. For example, in AES-T100 the Trojan leaks the private key through a covert channel using Code-Division Multiple Access (CDMA) communication. The Trojan employs a pseudo-random number generator (PRNG) to create a CDMA code sequence and XOR it with secret key bits. The modulated sequence is forwarded to a leakage circuit (LC) which leaks the sequence using the single output port. Our \textit{Intersect Analysis} detects the registers in the leakage circuit and also detects the malicious output ports through which the key is leaked. This Trojan is Type II Trojan meaning that the Trojan payload circuit is not part of the encryption algorithm. This Trojan is an always on Trojan and therefore, it does not have any triggering circuit.

AES-T200, AES-T700, AES-T900, AES-T1100 Trojan benchmarks use the same Trojan payload as AES-T100 to leak the secret key. All these Trojans are detected by our technique in the same way as the AES-T100 Trojan. Note that, the number of observe points and the malicious observe points are same for all these benchmarks. The reason is that these benchmark circuits use the same underlying  AES algorithm and have the same Trojan payload. The difference among these Trojans is how they are activated. AES-T700 is activated when it observes a specific triggering sequence whereas AES-T900 gets activated after a specific number of encryptions is performed. We detect these triggering conditions by analyzing the stimulus vector reported by our \textit{Confidentiality Verification} algorithm. The stimulus vector shows the condition for which an IFS violation occurs and these conditions represent the triggering sequence of the Trojan.

AES-T1100 Trojan's trigger circuit is composed of a finite state machine (FSM) and it triggers when a sequence of plaintexts is observed. The stimulus vector generated by our proposed framework will report the logic value of the Trojan state FFs for which the IFS violation occurs. However, it does not report the sequence of plaintexts for which the Trojan triggers. We apply the state machine extraction technique proposed in \cite{nahiyan2016avfsm} to retrieve the sequence of plaintexts which triggers the Trojan. This technique successfully extracts the triggering condition. Apart from AES-T1100, triggering circuit of AES-T2100 and AES-T1100(M) Trojan benchmarks also consists of an FSM. The trigger conditions of these Trojans were also successfully detected by \cite{nahiyan2016avfsm}. Note that, we identify the presence of FSM in the triggering circuit by identifying the state FF as discussed in Section \ref{sec:Trigger}.

AES-T1100(M) is the modified Trojan that we have presented in Section \ref{sec:Formal}. We modified the AES-T1100 Trojan triggering circuit by incorporating one more state (State 4 in Figure \ref{FSM}) and a 20 bit counter. Once the Trojan observes the four specific triggering sequence, it starts the counter and waits for $2^{20}$ cycles before triggering its payload (State 5 in Figure \ref{FSM}). This Trojan was also detected by our technique. Also, note that the time difference to detect AES-T1100 and AES-T1100(M) is very small, meaning that no matter how many additional states or cycles were introduced by the attacker, these Trojans will always be detected by our approach.

AES-T2000 and AES-T2100 Trojan benchmark have a different payload. The payload circuit of these Trojans consists of a shift register and it leaks the secret key through power side channel (PSC). These Trojans are also Type II Trojan and were detected by our intersect analysis. RSA-T100 and RSA-T300, on the other hand, are type I Trojans as they leak the key through valid ciphertext output. These Trojans were also detected by our technique as the propagation depth of these Trojan's leakage path is much less than the propagation depth of the valid paths.

To prove that our proposed framework can be applied to other designs, we implemented the payload circuit of the AES-T100 Trojan in DES and PRESENT encryption modules \cite{opencore}. We detected the Trojans in DES and PRESENT encryption modules as well.

\subsection{Integrity Verification}

We applied our \textit{Integrity Verification} algorithm to detect six different trust-hub Trojan benchmark circuits which violate the integrity policy and were able to detect all of them. Table \ref{tab:integrity} summarizes our results. Column 1 represents which Trojan benchmark was used for our verification while column 2 and 3 represent the Trojan payload and Trojan Trigger for each benchmark. Column 4 shows the Trojan types which was discussed in Section \ref{sec:Trojan}. Column 5 shows the number of control points for each asset bit while column 6 shows the malicious control points identified by the proposed approach. The selection of asset for \textit{Integrity Verification} is not as straightforward as the \textit{Confidentiality Verification}. For \textit{Integrity Verification}, we need to analyze the design and determine which asset an adversary may want to manipulate. For example, in a micro-processor an adversary may want to manipulate the program counter, instruction register, control and status registers (CSR). Therefore, all these registers need to be analyzed for integrity violation.

PIC-T100 Trojan manipulates the address of program memory, so that valid execution flow of the program is affected. Here, we make one bit of the program address register as the asset and apply our \textit{Integrity Verification}. The algorithm returns all the control points that can influence the asset. We know that the asset (program address register) should only be controlled by the program counter (PC) and based on this assumption we apply our \textit{Malicious Control Point Identification} technique. Our analysis returns that the asset can be influenced by a counter register which is not the PC. This proves the violation of integrity policy and presence of a malicious circuit. When we analyze the stimulus vector reported by out \textit{Integrity Verification}, we can find that the asset is influenced when the malicious counter reaches the value of $100$.

PIC-T200 and b19-T500 manipulate the instruction register of their corresponding micro-processor. Here, the asset is one bit of the instruction register. The instruction register should be controlled by the instruction cache. Our \textit{Integrity Verification} along with \textit{Malicious Control Point Identification} shows that the asset can be controlled by other malicious registers. The trigger condition for PIC-T200 was extracted by analyzing the stimulus vector. The trigger circuit of b19-T500 contains an FSM, and we determined the trigger condition by the FSM extraction technique proposed in \cite{nahiyan2016avfsm}.

s35932-T100 Trojan enables the scan enable signal of a part of one scan chain in the functional mode which allows an adversary to leak internal signal value. To detect this Trojan, we first identify the scan FFs which are not directly driven by scan enable signal. We then apply the \textit{Integrity Verification} to see which control points influence the scan enable port of these FFs. Our analysis shows when a certain counter value is reached, some internal registers take control of the scan enable port. This indicates a violation of integrity property as scan enable port of a scan FF should only be influenced by the scan enable signal. Our technique also successfully detected RS232-T500 Trojan which manipulates a control signal. 

The RSA-T400 benchmark contains a Trojan that replaces the key, which allows only the attacker to
decrypt the ciphertext. To detect this Trojan, we first apply \textit{Confidentiality Verification} algorithm to find which registers the key can propagate to. Next, we apply our \textit{Integrity Verification} in these registers. These key registers should only be influenced by the key and plaintext input. We applied our \textit{Malicious Control Point Identification} technique and found that certain internal registers can manipulate the key registers. We also extracted the trigger condition by analyzing the stimulus vector.

%----------------------- This is special for ITC --------------------------
% YOur copyright string might be differfent

\thispagestyle{fancy}
\fancyhead{}
\renewcommand{\headrulewidth}{0pt}
\fancyhf{}
\fancyfoot[L]{\large Paper 6.1}
\fancyfoot[C]{\large INTERNATIONAL TEST CONFERENCE}
\fancyfoot[R]{\large \thepage}

%----------------------- This is special for ITC --------------------------

\subsection{Comparison}
\label{Comparison}
In Section \ref{sec:Previous}, we have presented the limitations of the state of the art Trojan detection techniques. Here, we will experimentally validate that our proposed technique does not share these limitations. 

\noindent
{\bf Formal Methods:} Trojan detection based on formal methods, i.e., \cite{JV_1}, \cite{JV_2} are effective only to a limited number of clock cycles. We exploited this limitation to design AES-T1100(M) Trojan and experimentally verified that this Trojan cannot be detected by \cite{JV_1}, \cite{JV_2}. 

However, our proposed framework was able to detect this Trojan as shown in Table \ref{tab:confidentiality}. The reason is that our technique is not limited the number of clock cycles and therefore, no matter how many cycles are introduced by the attacker, this class of Trojans will always be detected by our approach.

Also, our technique tracks the actual asset propagation path and therefore, does not create \textit{False Positive Results} like \cite{JV_1} \cite{JV_2}.

\noindent
{\bf GLIFT:} GLIFT-based Trojan detection \cite{GLIFT} cannot distinguish between the valid key propagation path and the key leakage path caused by the Trojan. Therefore, Type I Trojans like RSA-T100 and RSA-T300 cannot be effectively detected by \cite{GLIFT}. However, our proposed framework was able to detect these Trojan (shown in Table \ref{tab:confidentiality}) by utilizing the propagation depth analysis.

Also, \cite{GLIFT} cannot detect Trojans, e.g., s35932-T100 which work in scan mode in DfT inserted netlist due to taint explosion. However, our technique can detect this Trojan (shown in Table \ref{tab:integrity}) because it can work with DfT inserted netlist.

\noindent
{\bf Jasper:} A SoC designer without the white box knowledge of an IP may not be able to utilize Jasper \cite{Jasper} tool to detect Trojan such as AES-T2000 and AES-T2100 which rely on the dynamic power consumption of internal registers to leak the key. However, our \textit{Malicious Observe Point Identification} technique can detect key leakage to these malicious registers. This feature allows us to detect them (shown in Table \ref{tab:confidentiality}).

\noindent
{\bf PCC:} Trojans like AES-T100 leak the key XORed with a PRNG value through a malicious output port. In \cite{Jin1}, the signal sensitivity of the key downgrades due to the XOR operation and the malicious output has a sensitivity value of $\{fix:0\}$. Meaning that this port does not leak any sensitive information and the Trojan would bypass the detection by \cite{Jin1}. Our framework was able to detect this Trojan (shown in Table \ref{tab:confidentiality}).

%\vspace{-1ex}
\section{Limitations of Proposed Approach}
\label{sec:discussion}

Our IFS verification framework relies on partial scan ATPG to identify all observe/control points to detect Trojan. However, it is possible that there exists observe/control points which are not detected by ATPG. This can happen if ATPG gives up when it takes too long to find patterns for asset propagation/activation. Also, as we increase the sequential depth of partial scan ATPG, it also increases the complexity to generate the required sequence of patterns. At some sequential depth, ATPG may fail to generate patterns to detect observe/control points. However, at such depth, an attacker would have little to gain by controlling and observing the control and observe points which were not found by the ATPG.

ATPG algorithm cannot work with latches and uncontrollable FFs as ATPG cannot control these components. Therefore, if a Trojan implementation contains latches or uncontrollable FFs, these observe/control points will not be detected by our proposed framework. However, it is highly unlikely that latches and uncontrollable FFs will be present in a properly implemented design. Moreover, our framework will issue a warning and will report the names along with the location of these components. We can then insert test points at these locations and analyze them using our technique.

%\vspace{-1ex}
\section{Conclusion}
\label{sec:concl}

In this paper, we have proposed a framework which detects violation of IFS policies caused by Trojans without the need of white-box knowledge of the IP. We have experimentally validated the effectiveness of our technique by detecting Trojan benchmarks from the trust-hub. We also compare our technique with the state-of-the-art Trojan detection techniques and validate that our proposed technique does not share the limitations of these techniques. Our proposed technique can identify IFS violation in a design. Here, we focused how to utilize it to detect hardware Trojans. However, this technique can also be used to detect IFS violations which are unintentionally introduced by design mistakes or CAD tools.

\section{Acknowledgment}

This work was supported in part by Semiconductor Research Corporation (SRC) and Cisco Systems, Inc. (Cisco). 

%----------------------- This is special for ITC --------------------------
% YOur copyright string might be differfent

\thispagestyle{fancy}
\fancyhead{}
\renewcommand{\headrulewidth}{0pt}
\fancyhf{}
\fancyfoot[L]{\large Paper 6.1}
\fancyfoot[C]{\large INTERNATIONAL TEST CONFERENCE}
\fancyfoot[R]{\large \thepage}

%----------------------- This is special for ITC --------------------------

%\vspace{-1ex}

\clearpage
\end{document}